\documentclass[preprint2]{aastex}
\usepackage{amsmath} 
\usepackage{color}

\shorttitle{electron drifts and KAW damping}
\shortauthors{Tong et al.}

\begin{document}
\title{Effects of electron drift on the collisionless damping of kinetic Alfv{\'e}n waves in the solar wind}

\author{Yuguang Tong\altaffilmark{1, 2}, Stuart D. Bale\altaffilmark{1, 2},
Christopher H. K. Chen\altaffilmark{3, 1},
Chadi S. Salem\altaffilmark{1} and
Daniel Verscharen\altaffilmark{4}
}

\altaffiltext{1}{Space Sciences Laboratory, University of California, Berkeley, CA 94720, USA
bale@ssl.berkeley.edu}
\altaffiltext{2}{Department of Physics, University of California, Berkeley, CA 94720, USA ygtong@berkeley.edu}
\altaffiltext{3}{Department of Physics, Imperial College London, London SW7 2AZ, UK}
\altaffiltext{4}{Space Science Center, University of New Hampshire, Durham, NH 03824, USA}
\begin{abstract}
The collisionless dissipation of anisotropic Alfv{\'e}nic turbulence is a promising candidate to solve the solar wind heating problem. Extensive studies examined the kinetic properties of Alfv{\'e}n waves in simple Maxwellian or bi-Maxwellian plasmas. However, the observed electron velocity distribution functions in the solar wind are more complex. In this study, we analyze the properties of kinetic Alfv\'en waves in a plasma with two drifting electron populations. We numerically solve the linearized Maxwell-Vlasov equations and find that the damping rate and the proton-electron energy partition for kinetic Alfv{\'e}n waves are significantly modified in such plasmas, compared to plasmas without electron drifts. We suggest that electron drift is an important factor to take into account when considering the dissipation of Alfv{\'e}nic turbulence in the solar wind or other $\beta \sim 1$ astrophysical plasmas.
\vspace{0.5in}
\end{abstract}

\section{Introduction}
One major question in space physics is how collisionless plasmas in the extended corona and in the solar wind are heated. Observations in the solar wind reveal that the fluctuations of the electric and magnetic fields show a turbulent spectrum similar to the power-law spectrum of fluid turbulence as described by Kolmogorov \citep{Tu:1995a, Bale:2005a, Bruno:2013a, Alexandrova:2013a}. The  turbulence in the solar wind shows mainly Alfv{\'e}nic polarization \citep{Belcher:1971a} and becomes more anisotropic during the cascade to higher wavenumbers $k$ \citep{Goldreich:1995b, Horbury:2008a, chen:2010a}, resulting in $k_{\perp}\gg k_{\parallel}$ at short wavelengths. At the proton scale, the turbulence is thought to transition into a kinetic Alfv{\'e}n wave (KAW) cascade \citep{Schekochihin:2009a, Sahraoui:2010a, Howes:2011c, Salem:2012a, Boldyrev:2012a, Chen:2013a}, which gradually dissipates energy to the particles \citep{Leamon:1999a, Howes:2011c}.

Despite the nonlinear nature of turbulence, linear theory has been used in studies of solar wind turbulence (see, for instance, \citet{Howes:2014a} for review). These studies analyze turbulent fluctuations by means of the linear propagation and damping characteristics of KAWs to help to understand the dissipation of the turbulence and the heating of the solar wind. The vast majority of these studies rely on representing the velocity distribution functions (VDFs) by an isotropic or bi-Maxwellian background (e.g., \cite{Quataert:1998a, Leamon:1999a, Cranmer:2003a}). In this work, we investigate how more realistic VDFs modify the damping of KAWs. In the solar wind, both ions and electrons can be modeled by several populations drifting with respect to each other.  Recent studies show that the differential flow between ions affects ion heating by both ion-cyclotron waves \citep{Kasper:2013a} and low-frequency KAWs \citep{Chandran:2013a}.

The solar-wind electron VDF can be modeled as a superposition of three electron populations: a cool and dense ``core'', a hot and less dense ``halo'' \citep{Feldman:1975a} and a beam-like one-sided ``strahl" \citep{Maksimovic:2005a}. In the proton frame, core (halo and strahl) electrons drift sunward (anti-sunward) along the background magnetic field. Empirically, the solar wind fulfills quasi-neutrality and the zero-current condition \citep{Feldman:1975a, Pulupa:2014a}.  The uncertainty in the fitting parameters for electron core properties is smaller than the uncertainty in the fitting parameters for halo and strahl properties. A drifting bi-Maxwellian population models core electrons very well. In particular, the core electron bulk drift speed is usually 
comparable to or larger than the Alfv{\'e}n speed and shows a clear statistical dependence on collisional age \citep{Pulupa:2014a}. In the extrapolated asymptotic limit of no collisions, the core drift is as large as three to four times the Alfv{\'e}n speed.

In this letter,  we examine how the drift between electron core and halo affects the linear damping of KAWs.  We model the solar wind electrons by a superposition of a Maxwellian core and a Maxwellian halo drifting against each other in the proton frame.  Choosing a set of parameters typical for the solar wind, we solve the full hot-plasma dispersion relation of KAWs in the framework of Vlasov-Maxwell theory and identify the contribution to wave damping from each plasma component. We show that, in the linear approximation, electron drifts can lead to significant variations in both the KAW damping rates and the relative  energy transfer from the waves to protons and  electrons, the latter of which is another important unsolved problem in turbulent plasma heating. 

\section{Theory and method}
\label{sec: theory&method}
\begin{figure}
	\centering
	\includegraphics[width=0.8\linewidth]{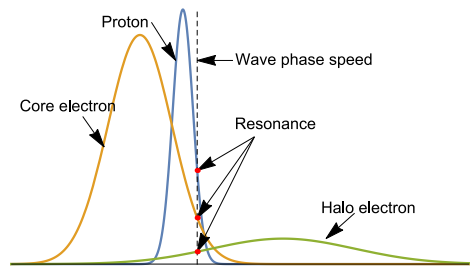}
	\caption{Schematic diagram for  the   plasma considered in this letter.  We annotate the parallel  wave phase speed $v_{\mathrm{res}}=\omega_r/k_{\parallel}$ for  our   resonance analysis in Section \ref{sec:results}.}
	\label{fig:cartoon}
\end{figure}

Motivated by observed solar wind electron VDFs,  we consider a collisionless, homogeneous, and warm proton-electron plasma in a constant uniform background magnetic field. The proton temperature and density are denoted as $T_p$ and $n_p$; the electron distribution consists of two shifted Maxwellian populations, a cool ($T_c$) ``core'' and a hot ($T_h$) ``halo.'' As an initial step to study the effects of core electron drifts on resonant damping, we use isotropic temperatures for all plasma components. Both electron populations drift along the background magnetic field $\mathbf{B}_0=B_0\hat{\mathbf{z}}$ with $\mathbf{v}_c=v_c\hat{\mathbf{z}}$ and $\mathbf{v}_h=v_h\hat{\mathbf{z}}$ in the proton frame. The halo electron drift adjusts to the core electron drift to guarantee the absence of net currents in the plasma: $v_h= - v_c n_c/n_h$, where $n_c$ and $n_h$ denote core and halo electron densities. Figure \ref{fig:cartoon} shows the plasma described above schematically.

Under the assumption of strong anisotropy in kinetic scale fluctuations in the solar wind, we consider KAWs with wavevectors at a large angle with respect to the magnetic field. We use constant values for the angle between $\mathbf{k}$ (wavenumber vector) and $\mathbf{B}_0$ ($89^\circ$ or $91^\circ$), plasma beta,  the electron-core-to-proton temperature ratio and the core-to-halo density ratio, so that the only free parameters are $\mathbf{k}$, $\mathbf{v}_c$ and $\mathrm{sign}(\mathbf{k}\cdot \mathbf{B}_0)$. In particular, we allow $\mathbf{B}_0$ to be either parallel or anti-parallel to $\mathbf{v}_c$, and $\mathbf{k}$ has a finite component along $\mathbf{B}_0$. In principle, each configuration $(\mathbf{v}_c, \mathbf{k}, \mathbf{B}_0)$ requires a separate treatment; however, changing the direction of $\mathbf{B}_0$ only inverts the direction of propagation and does not produce a new dispersion relation. Therefore, we introduce  the dimensionless core electron drift $\delta v_c \equiv |v_c/v_A|\mathrm{sign}(\mathbf{k}\cdot \mathbf{v}_c)$,  where $v_A\equiv \sqrt{B_0^2/4\pi n_p m_p}$ is the proton Alfv{\'e}n speed. The values of the remaining parameters are given in Table \ref{tab: plasma parameters}. 

\begin{table}
\centering
\begin{tabular}{l |c}
\tableline 
Quantity &  Value \\ 
\tableline 
\tableline
Proton plasma beta $\beta_p$ & $0.4$ \\ 
Temperature ratio  $T_c/T_p$ & $2$ \\ 
$T_h/T_p$& $10$ \\ 
Density ratio $n_c/n_h$ & $9$ \\ 
Proton bulk drift $v_p/v_A$ & $0$ \\ 
Angle between $\mathbf{k}$,  $\mathbf{B}_0$ & $89^\circ$ or $91^\circ$ \\
\tableline
\end{tabular} 
\caption{Values of fixed plasma parameters in our study.}
\label{tab: plasma parameters}
\end{table}

We study the dispersion relations of KAWs by numerically solving the full set of the linear Maxwell-Vlasov equations. Linear Maxwell-Vlasov theory has been described in detail in the literature   (see, for instance, \cite{Swanson:1989a} and \cite{Stix:1992a}). We summarize the relevant results here. Linear plasma waves are eigenmodes of the wave equation:
\begin{equation}
\mathbf{k}\times\left(\mathbf{k}\times\mathbf{E}\right)+\frac{\omega^2}{c^2}\boldsymbol{\epsilon}\cdot\mathbf{E}=0,
\end{equation}
where $\mathbf{E}$ is the fluctuating electric field in Fourier space. The dielectric tensor $\boldsymbol{\epsilon}$ incorporates contributions from each plasma component:  $\boldsymbol{\epsilon}=\boldsymbol{1}+\Sigma_s\boldsymbol{\chi}_s $,  where $\boldsymbol{\chi_s}$ is the susceptibility tensor for species $s$. We obtain $\boldsymbol{\chi_s}$  from the linearized  and Fourier-transformed Vlasov equation and Maxwell's equations.

To study the contribution to wave damping (growth) from each component, we calculate the relative particle heating rates as \citep{Stix:1992a} 
\begin{equation}
P_s\equiv\mathbf{E}^\ast\cdot\boldsymbol{\chi}^a_s\arrowvert_{\omega=\omega_r}\cdot\mathbf{E}/4W.
\label{eq:ps}
\end{equation}
where $\boldsymbol{\chi}^a_s\equiv \left(\boldsymbol{\chi_s}-\boldsymbol{\chi}_s^\dagger\right)/2i$ is the anti-Hermitian part of $\boldsymbol{\chi_s}$, $\omega_r\equiv\Re\left(\omega\right)$ and $W$ is the wave energy.

In the  weak damping limit, $P_s$ gives the fraction of wave energy damped by plasma component $s$ during  one   wave period. The total heating rate $P_{total} \equiv \sum_s P_s$  corresponds to the   wave energy dissipation rate directly calculated from the damping rate, i.e., $1-e^{2\omega_iT}$, where $\omega_i\equiv\Im\left(\omega\right)$ and $T\equiv2\pi/\omega_r$ is the wave period,  as long as $\omega_i$ is small compared to $\omega_r$. In this letter, we focus on the linear damping of KAWs. Electron drifts may, however, alter wave--particle interactions to such a degree that the corresponding waves become unstable, which we note for the sake of completeness. Notice that $P_s$ may become negative, corresponding to cooling of species $s$.

\section{Results}
\label{sec:results}
\begin{figure}
	\centering
	\includegraphics[width=0.8\linewidth]{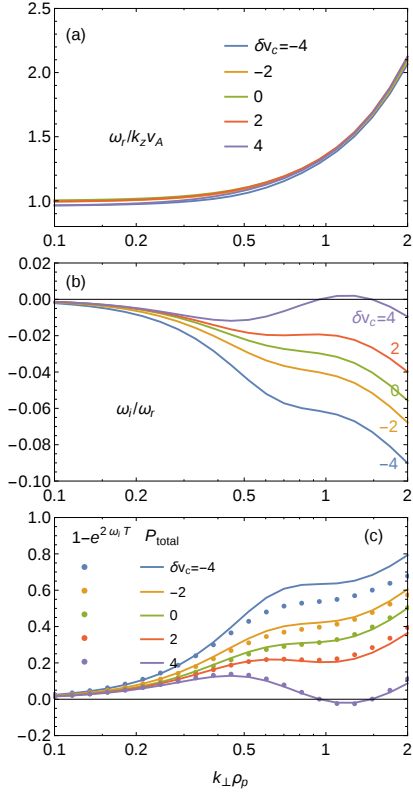}
	\caption{(a)-(b): Dispersion relation of KAWs in plasmas for  different   $\delta v_c$. $\omega_r$ and $\omega_i$ are  the  real and imaginary parts of the wave frequency. $\rho_p\equiv v_{Tp}/\Omega_p$ is  the proton gyro-radius, $v_{Tp}\equiv \sqrt{2k_BT_p/m_p}$ is the proton thermal speed, and $\Omega_p\equiv eB_0/m_pc$ is the proton gyro-frequency.
    (c): Total particle heating rates ($P_{total}$, solid lines) and wave energy dissipation rates ($1-e^{2\omega_iT}$, dots) as functions of wavenumber.
    }	
	\label{fig:panel1}
\end{figure}

Figures \ref{fig:panel1}(a)-(b) show dispersion relations of KAWs in plasmas with $\delta v_c \in [-4, 4]$.  Panel (a) demonstrates that electron drifts have little effect on the real part of  the  KAW frequencies, consistent with \cite{Gary:1975b}. Indeed,  the  protons rather than  the   electrons  mainly determine the dynamics of low-frequency  KAWs. Panel (b) compares damping rates of KAWs in plasmas with different electron drifts. Negative (positive) $\delta v_c$  lead to an enhancement (reduction) of the   KAW damping  at   all wavenumbers. Panel (b) shows that $\delta v_c \sim -4$  leads to an increase in the   damping rate by $\sim50\%$  compared to the case without electron drifts. For sufficiently large positive $\delta v_c$, KAWs become unstable  in  certain  wavenumber  ranges. This instability is an example of a ``heat flux instability" \citep{Gary:1975b}. The two-population electrons in our model introduce an electron heat flux (third moment of the electron VDF), which may provide energy to drive certain wave modes (e.g., whistler, Alfv{\'e}n, and magnetosonic) unstable \citep{Gary:1975b, Gary:1998a, Gary:1999a}. These heat flux instabilities, in return, regulate the electron heat flux and hence the electron drifts. We note that there is no conflict between the damping and growth of KAWs and other wave modes. For instance, parallel whistler waves, whose instability threshold is much lower than that of the Alfv\'enic instability, are unstable even if $\delta v_c>-4$. Therefore, the Alfv\'enic instability cannot regulate the electron heat flux in the solar wind and is more of academic interest under typical solar-wind conditions since the instability with the lowest threshold constrains the electron heat flux once triggered.
Figure \ref{fig:panel1}(c) compares the total particle heating rate (solid) with the wave energy dissipation rate (dotted). When $1-e^{2\omega_iT}\lesssim 0.5$, \textit{i.e.}, the wave retains more than half of its energy after a wave period, $P_{total}\approx1-e^{2\omega_iT}$. Hence we show that $P_s$ is a good measure of energy flow as long as damping is weak. 

\begin{figure*}
	\centering
	\includegraphics[width=0.8\linewidth]{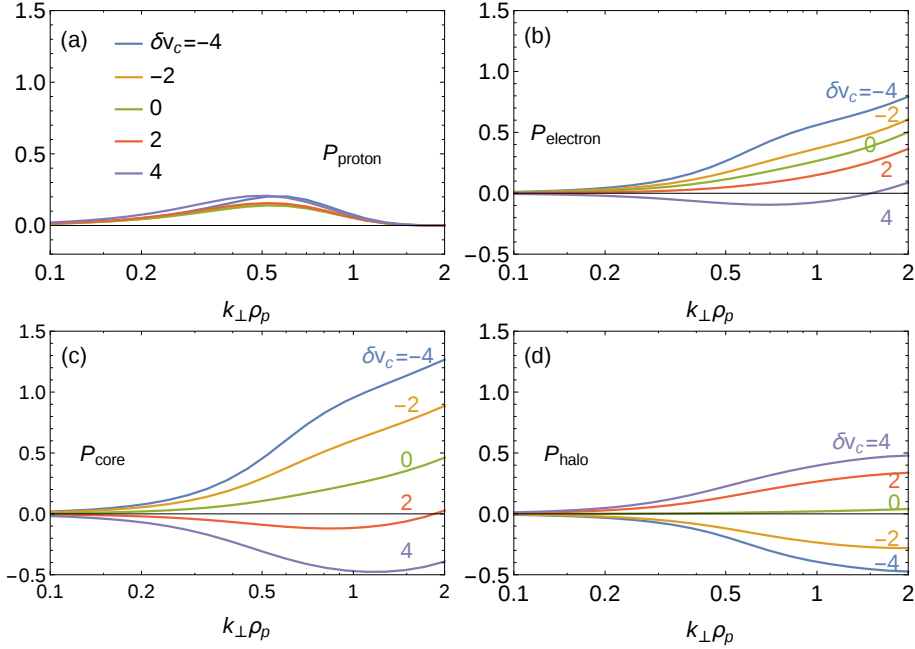}
	\caption{Dependence of $P_{proton}$ (a), $P_{electron}$ (b), $P_{core}$(c) and $P_{halo}$ (d) on $k_\perp\rho_p$ in plasmas with varying $\delta v_c$.}
	\label{fig: p_s vs k }
\end{figure*}

Figures \ref{fig: p_s vs k } (a)-(d) present heating rates for protons,  all  electrons (core + halo),  as well as separately for   core electrons and halo electrons. Recall that positive $P_s$ indicates heating for species $s$, and that negative $P_s$ indicates cooling. Panel (a) shows that protons are always heated, although  electron drifts can slightly modify  the values of the proton heating rate.  On the other hand, panels (b)-(d) show that an electron population can  experience both cooling and heating depending on $\delta v_c$. While both core and halo are heated in the absence of bulk drifts, sufficiently large positive $\delta v_c$ lead to core cooling and at the same time significantly enhance halo heating. Negative $\delta v_c$ have the opposite effect on electron heating.  In general, electron drifts have a larger effect on electron heating than on proton heating. Electron drifts modify the damping  rates   of KAWs by changing the  efficiency   of wave--particle interactions for  the different   electron populations. 

\begin{figure}
	\centering
	\includegraphics[width=0.8\linewidth]{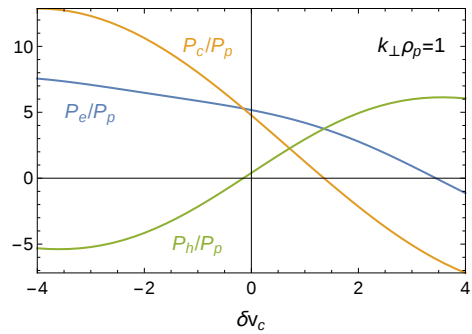}
	\caption{Ratio of energy partition between electrons and protons as a function of electron drift.}
	\label{fig: energy_partition}
\end{figure}

Figure \ref{fig: energy_partition} shows how  the  energy partition depends on electron drifts at $k_{\perp}\rho_p=1$. $P_c/P_p$ and $P_h/P_p$ vary wildly with $\delta v_c$. However,  since electron drifts affect core heating and halo heating in opposite ways, the total electron heating and  consequently the   electron-proton energy partition  reveal   a more moderate dependence on electron drifts.  Nevertheless, a core drift of   $\delta v_c \sim -4$ increases $P_e/P_p$ by $\sim25\%$. Without electron drifts, electron heating dominates over proton heating in hydrogen plasmas at $k_{\perp}\rho_p\sim 1$ for $\beta\sim 1$ \citep{Quataert:1998a}. We see in Figure \ref{fig: energy_partition} that positive $\delta v_c$ significantly reduces electron-to-proton heating ratio. At $\delta v_c\sim 3$, electron heating becomes negligible compared to proton heating. We note at this point that  the  dependence of energy partition on electron drifts is qualitatively the same  at other wavenumbers, which can be inferred from Figure \ref{fig: p_s vs k }.

 We interpret  the  dependence of wave damping and particle heating on electron drifts  in terms of a simplified resonance analysis. Since $\omega_r$ is almost independent of electron drifts, the proton heating rate shows only a small variation. However, core and halo electron bulk drifts  significantly change the value and the gradient of the eVDF at the resonance speed,
\begin{equation}
v_{\mathrm{res}}\equiv\frac{\omega_{r}-n\Omega_e}{k_{\parallel}},
\end{equation}
where $\Omega_e$ is the electron gyro-frequency and $n$ is an integer depending on the polarization properties of the wave mode \citep{Marsch:2006a}. 

In our case, the most relevant resonance for wave--particle interactions with KAWs is the Landau resonance with $n=0$. Electrons with the (field-parallel) speed $v_{\mathrm{res}}$ interact resonantly with the corresponding wave mode. We illustrate   the case for negative $\delta v_c$ schematically in Figure \ref{fig:cartoon}. 

 The gradient of the distribution functions at the resonance speed determines if the particle species gains or loses energy, i.e., the signs of $P_p$, $P_c$, and $P_h$. In the example shown in Figure~\ref{fig:cartoon},  the resonance occurs where  the  halo VDF has  a  positive  gradient, and hence  the  KAW  removes   energy from  the   halo electrons,  leading to a   negative  value for  $P_h$. In contrast, core electrons and protons have resonances where  the gradients of their  VDFs are negative,  leading to positive values for    $P_c$ and $P_p$. Similarly, the dependence of $P_s$ on $\delta v_c$ can also be inferred from  the  positions of  the   resonance points  with respect to the VDFs. However, a qualitative resonance analysis  cannot provide a thorough comparison across the plasma components. We find that $P_s$, defined in Eq. (\ref{eq:ps}), gives a better understanding of  the  energy  transfer   between waves and different plasma components.

It is worth noting that electron drifts seem to provide a collisionless energy coupling between core and halo electrons (see the case with $\delta v_c=-2$ for instance). Figure \ref{fig:panel1}(c) shows that $1-e^{2\omega_iT} \approx P_c + P_h + P_p$. With $P_c, P_p>0 $ and $P_h<0$ (from Figure \ref{fig: p_s vs k }), $\left(1-e^{2\omega_iT}\right) + (-P_h)\approx P_c +P_p $ , meaning that energy flows from damped waves and halo electrons into core electrons and protons. A similar energy transport among electron components was suggested by \citet{Gary:1998b}. According to this study, the Alfv\'enic heat-flux instability transfers energy from the drifting halo into the electron core through the Landau-resonant absorption of the waves, leading to a lower limit for $\beta_c$.

\section{Discussion and conclusion}
For  the sake of  clarity, we present results for a single propagation angle in this letter. A careful analysis shows that  other oblique angles do not lead to any  qualitative differences. Given a fixed KAW power spectrum,  we anticipate that the  presence of electron drifts changes  the  wave-energy dissipation rate, particle heating rates, and  the  proton-electron energy partition ratio  significantly. 

In the context of  the  solar wind, since core electrons are nearly always observed to drift sunward in  the   proton frame, positive (negative) $\delta v_c$ correspond to KAWs propagating sunward (anti-sunward).  Therefore, anti-sunward KAWs experience stronger damping  in the presence of electron drifts and lead to a stronger core heating, halo electron cooling, and a stronger total electron heating. Since the total energy flux typically points anti-sunward in the solar wind, we suggest that electron drifts lead to stronger electron heating than  expected from previous calculations \citep{Quataert:1998a,  Leamon:1999a}.  An interesting corollary of this work is that, since the core electron drift speed depends on the collisionality of the solar wind \citep{Pulupa:2014a}, the heating rate as calculated here may therfore also indirectly depend on solar wind Couloumb collisions. 

Our model includes relative drifts in the electron VDF, while it ignores several additional features of solar wind electrons, namely the presence of superthermal electrons, one-sided beams (strahl)  and temperature anisotropies.  By using a Maxwellian halo, we ignore high energy tails in the electron VDF. However, this assumption does not significantly affect the gradients of the electron distribution function at $v_{\mathrm{res}}\sim v_{A}$ which mainly determine the damping. We can apply a similar argument to strahl electrons, which also occupy a different region in the velocity space. Regarding temperature anisotropies, \cite{Gary:1975b} studied the dependence of the core drift threshold for the Alfv\'enic instability on electron temperature anisotropy and found that the dependence is weak in plasmas with $\beta_p<0.25$. We expect a similar weak relation in our case, since Landau-resonant instabilities show such a weak dependence in general \citep{Verscharen:2013}, since the strength of this type of resonant interaction is determined by the parallel gradients of the electron distribution function. If the parallel temperature is kept constant, the introduction of temperature anisotropy only changes the perpendicular gradients of the VDF and therefore does not significantly alter the wave--particle interaction.

In this letter, we choose a fiducial set of representative solar wind plasma parameters to demonstrate the effects of electron drifts on the damping of KAWs. In order to fully account for all of the relevant effects in the solar wind, it is important to conduct a full scan of the corresponding parameter space. This endeavor is beyond the scope of this work and will be presented in a future paper.

Our work shows that, despite our limiting assumption of a superposition of Maxwellians to represent the electron distribution function, its fine-structure has a strong influence on the propagation and damping properties of KAWs. Therefore, our work is of relevance in the broader context of all collisionless astrophysical plasmas in which non-Maxwellian electron distributions can develop and persist. Considering that most astrophysical plasmas are in a turbulent state and that KAW-turbulence is believed to be the dominant type of plasma turbulence on small scales, our work suggests that the effects of non-Maxwellian electron distributions be carefully accounted for in studies of collisionless astrophysical plasmas in general.

We are grateful to Christopher C. Chaston, Marc P. Pulupa, Eliot Quataert and Kristopher G. Klein for helpful discussions. Y. Tong is supported by NASA grant APL-975268 and Charles K. Kao Scholarship. S. D. Bale is supported by NASA grant APL-975268. C. H. K. Chen is supported by an Imperial College Junior Research Fellowship. C. Salem is supported by NASA grant NNX14AC07G. D. Verscharen is supported by NASA grant NNX12AB27G.


\begin{thebibliography}{}
\expandafter\ifx\csname \natexlab\endcsname\relax\def\natexlab#1{#1}\fi

\bibitem[{{Alexandrova} {et~al.}(2013){Alexandrova}, {Chen}, {Sorriso-Valvo},
  {Horbury}, \& {Bale}}]{Alexandrova:2013a}
{Alexandrova}, O., {Chen}, C.~H.~K., {Sorriso-Valvo}, L., {Horbury}, T.~S., \&
  {Bale}, S.~D. 2013, \ssr, 178, 101

\bibitem[{{Bale} {et~al.}(2005){Bale}, {Kellogg}, {Mozer}, {Horbury}, \&
  {Reme}}]{Bale:2005a}
{Bale}, S.~D., {Kellogg}, P.~J., {Mozer}, F.~S., {Horbury}, T.~S., \& {Reme},
  H. 2005, \prl, 94, 215002

\bibitem[{{Bale} {et~al.}(2013){Bale}, {Pulupa
}, {Salem}, {Chen}, \&
  {Quataert}}]{Bale:2013a}
{Bale}, S.~D., {Pulupa}, M., {Salem}, C., {Chen}, C.~H.~K., \& {Quataert}, E.
  2013, \apjl, 769, L22

\bibitem[{{Belcher} \& {Davis}(1971)}]{Belcher:1971a}
{Belcher}, J.~W., \& {Davis}, Jr., L. 1971, \jgr, 76, 3534

\bibitem[{{Boldyrev} \& {Perez}(2012)}]{Boldyrev:2012a}
{Boldyrev}, S., \& {Perez}, J.~C. 2012, \apjl, 758, L44

\bibitem[{{Bruno} \& {Carbone}(2013)}]{Bruno:2013a}
{Bruno}, R., \& {Carbone}, V. 2013, Living Reviews in Solar Physics, 10, 2

\bibitem[{Chandran {et~al.}(2013)Chandran, Verscharen, Quataert, Kasper,
  Isenberg, \& Bourouaine}]{Chandran:2013a}
Chandran, B. D.~G., Verscharen, D., Quataert, E., {et~al.} 2013, \apj, 776, 45

\bibitem[{{Chen} {et~al.}(2013){Chen}, {Boldyrev}, {Xia}, \&
  {Perez}}]{Chen:2013a}
{Chen}, C.~H.~K., {Boldyrev}, S., {Xia}, Q., \& {Perez}, J.~C. 2013, \prl, 110,
  225002

\bibitem[{{Chen} {et~al.}(2010){Chen}, {Horbury}, {Schekochihin}, {Wicks},
  {Alexandrova}, \& {Mitchell}}]{chen:2010a}
{Chen}, C.~H.~K., {Horbury}, T.~S., {Schekochihin}, A.~A., {et~al.} 2010, \prl,
  104, 255002

\bibitem[{{Cranmer} \& {van Ballegooijen}(2003)}]{Cranmer:2003a}
{Cranmer}, S.~R., \& {van Ballegooijen}, A.~A. 2003, \apj, 594, 573

\bibitem[{{Feldman} {et~al.}(1975){Feldman}, {Asbridge}, {Bame}, {Montgomery},
  \& {Gary}}]{Feldman:1975a}
{Feldman}, W.~C., {Asbridge}, J.~R., {Bame}, S.~J., {Montgomery}, M.~D., \&
  {Gary}, S.~P. 1975, \jgr, 80, 4181

\bibitem[{{Gary} {et~al.}(1975){Gary}, {Feldman}, {Forslund}, \&
  {Montgomery}}]{Gary:1975b}
{Gary}, S.~P., {Feldman}, W.~C., {Forslund}, D.~W., \& {Montgomery}, M.~D.
  1975, \jgr, 80, 4197
  
  \bibitem[{{Gary} {et~al.}(1998{\natexlab{a}}){Gary}, {Li}, {O'Rourke}, \&
  {Winske}}]{Gary:1998a}
{Gary}, S.~P., {Li}, H., {O'Rourke}, S., \& {Winske}, D. 1998{\natexlab{a}},
  \jgr, 103, 14567

\bibitem[{{Gary} {et~al.}(1998{\natexlab{b}}){Gary}, {Newbury}, \&
  {Goldstein}}]{Gary:1998b}
{Gary}, S.~P., {Newbury}, J.~A., \& {Goldstein}, B.~E. 1998{\natexlab{b}},
  \jgr, 103, 14559
  
\bibitem[{{Gary} {et~al.}(1999){Gary}, {Skoug}, \& {Daughton}}]{Gary:1999a}
{Gary}, S.~P., {Skoug}, R.~M., \& {Daughton}, W. 1999, Physics of Plasmas, 6,
  2607
 

\bibitem[{{Goldreich} \& {Sridhar}(1995)}]{Goldreich:1995b}
{Goldreich}, P., \& {Sridhar}, S. 1995, \apj, 438, 763

\bibitem[{{Horbury} {et~al.}(2008){Horbury}, {Forman}, \&
  {Oughton}}]{Horbury:2008a}
{Horbury}, T.~S., {Forman}, M., \& {Oughton}, S. 2008, \prl, 101, 175005

\bibitem[{{Howes} {et~al.}(2014){Howes}, {Klein}, \& {TenBarge}}]{Howes:2014a}
{Howes}, G.~G., {Klein}, K.~G., \& {TenBarge}, J.~M. 2014, arXiv:1404.2913

\bibitem[{{Howes} {et~al.}(2011){Howes}, {Tenbarge}, {Dorland}, {Quataert},
  {Schekochihin}, {Numata}, \& {Tatsuno}}]{Howes:2011c}
{Howes}, G.~G., {Tenbarge}, J.~M., {Dorland}, W., {et~al.} 2011, \prl, 107,
  035004

\bibitem[{{Kasper} {et~al.}(2013){Kasper}, {Maruca}, {Stevens}, \&
  {Zaslavsky}}]{Kasper:2013a}
{Kasper}, J.~C., {Maruca}, B.~A., {Stevens}, M.~L., \& {Zaslavsky}, A. 2013,
  \prl, 110, 091102

\bibitem[{{Leamon} {et~al.}(1999){Leamon}, {Smith}, {Ness}, \&
  {Wong}}]{Leamon:1999a}
{Leamon}, R.~J., {Smith}, C.~W., {Ness}, N.~F., \& {Wong}, H.~K. 1999, \jgr,
  104, 22331

\bibitem[{{Maksimovic} {et~al.}(2005){Maksimovic}, {Zouganelis}, {Chaufray},
  {Issautier}, {Scime}, {Littleton}, {Marsch}, {McComas}, {Salem}, {Lin}, \&
  {Elliott}}]{Maksimovic:2005a}
{Maksimovic}, M., {Zouganelis}, I., {Chaufray}, J.-Y., {et~al.} 2005, \jgr,
  110, 9104

\bibitem[{{Marsch}(2006)}]{Marsch:2006a}
{Marsch}, E. 2006, Living Reviews in Solar Physics, 3, 1

\bibitem[{{Pulupa} {et~al.}(2014){Pulupa}, {Bale}, {Salem}, \&
  {Horaites}}]{Pulupa:2014a}
{Pulupa}, M.~P., {Bale}, S.~D., {Salem}, C., \& {Horaites}, K. 2014, \jgr, 119,
  647

\bibitem[{{Quataert}(1998)}]{Quataert:1998a}
{Quataert}, E. 1998, \apj, 500, 978

\bibitem[{{Sahraoui} {et~al.}(2010){Sahraoui}, {Goldstein}, {Belmont}, {Canu},
  \& {Rezeau}}]{Sahraoui:2010a}
{Sahraoui}, F., {Goldstein}, M.~L., {Belmont}, G., {Canu}, P., \& {Rezeau}, L.
  2010, \prl, 105, 131101

\bibitem[{{Salem} {et~al.}(2012){Salem}, {Howes}, {Sundkvist}, {Bale},
  {Chaston}, {Chen}, \& {Mozer}}]{Salem:2012a}
{Salem}, C.~S., {Howes}, G.~G., {Sundkvist}, D., {et~al.} 2012, \apjl, 745, L9

\bibitem[{{Schekochihin} {et~al.}(2009){Schekochihin}, {Cowley}, {Dorland},
  {Hammett}, {Howes}, {Quataert}, \& {Tatsuno}}]{Schekochihin:2009a}
{Schekochihin}, A.~A., {Cowley}, S.~C., {Dorland}, W., {et~al.} 2009, \apjs,
  182, 310

\bibitem[{{Stix}(1992)}]{Stix:1992a}
{Stix}, T.~H. 1992, {Waves in plasmas} (American Institute of Physics)

\bibitem[{{Swanson}(1989)}]{Swanson:1989a}
{Swanson}, D.~G. 1989, {Plasma waves} (Academic Press)

\bibitem[{{Tu} \& {Marsch}(1995)}]{Tu:1995a}
{Tu}, C.-Y., \& {Marsch}, E. 1995, \ssr, 73, 1

\bibitem[{Verscharen} \& {Chandran} (2013)]{Verscharen:2013}{Verscharen}, D. \& {Chandran}, B.~D.~G., 2013, \apj, 764, 88

\end{thebibliography}
\end{document}